\documentclass[twocolumn,english,aps,prl,superscriptaddress,amsmath,floatfix,showpacs]{revtex4}
\usepackage[T1]{fontenc}
\usepackage[latin9]{inputenc}
\usepackage{color}
\usepackage{graphicx}
\usepackage{amssymb}

\makeatletter
\@ifundefined{textcolor}{}
{%
 \definecolor{BLACK}{gray}{0}
 \definecolor{WHITE}{gray}{1}
 \definecolor{RED}{rgb}{1,0,0}
 \definecolor{GREEN}{rgb}{0,1,0}
 \definecolor{BLUE}{rgb}{0,0,1}
 \definecolor{CYAN}{cmyk}{1,0,0,0}
 \definecolor{MAGENTA}{cmyk}{0,1,0,0}
 \definecolor{YELLOW}{cmyk}{0,0,1,0}
 }

\usepackage{graphics}\usepackage{epsfig}\@ifundefined{definecolor}
 {\usepackage{color}}{}

\def\G0W0{$\mathrm G^\circ\mathrm W^\circ$}

\makeatother

\usepackage{babel}

\begin{document}

\title{GW quasi-particle spectra from occupied states only }

\author{P. Umari}

\affiliation{INFM-CNR DEMOCRITOS Theory@Elettra group, c/o Sincrotrone Trieste,
Area Science Park, I-34012 Basovizza, Trieste, Italy }

\author{Geoffrey Stenuit}

\affiliation{INFM-CNR DEMOCRITOS Theory@Elettra group, c/o Sincrotrone Trieste,
Area Science Park, I-34012 Basovizza, Trieste, Italy }

\author{Stefano Baroni}

\affiliation{SISSA -- Scuola Internazionale Superiore di Studi Avanzati, via Beirut
2-4, I-34151 Trieste Grignano, Italy }

\affiliation{INFM-CNR DEMOCRITOS Theory@Elettra group, c/o Sincrotrone Trieste,
Area Science Park, I-34012 Basovizza, Trieste, Italy }

\date{\today}
\begin{abstract}
We introduce a method that allows for the calculation of quasi-particle
spectra in the GW approximation, yet avoiding any explicit reference
to empty one-electron states. This is achieved by expressing the irreducible
polarizability operator and the self-energy operator through a set
of linear response equations, which are solved using a Lanczos-chain
algorithm. We first validate our approach by calculating the vertical
ionization energies of the benzene molecule and then show its potential
by addressing the spectrum of a large molecule such as free-base tetraphenylporphyrin. 
\end{abstract}


\maketitle

In spite of the formidable success met over the past forty years in
the simulation of materials, based on electronic-structure theory
\cite{martin-book}, density-functional theory (DFT) \cite{dft} is
essentially limited to ground-state properties and its time-dependent
extension \cite{tddft} still displays conceptual and practical difficulties.
Many-body perturbation theory (MBPT) \cite{hl69}, in turn, provides
a general, though awkward, framework for simulating electronic excitation
processes in materials. The most elementary such process is the removal/addition
of an electron from a system originally in its ground state. These
processes are accessible to direct/inverse photo-emission spectroscopies
and can be described theoretically in terms of \emph{quasi-particle}
(QP) spectra \cite{hl69}. A numerically viable approach to QP energy
levels (known as the GW approximation, GWA) was introduced in the
60's \cite{h65,ag98}, but it took two decades for a realistic application
of it to appear \cite{hl85}, and still now the routine application
of MBPT to the simulation of materials is plagued by severe numerical
difficulties, which have limited so far these applications to models
of a few handfuls of nonequivalent atoms, at most. The two main such
difficulties are the necessity to calculate and manipulate large matrices
representing the charge response of the system (electron polarizabilities
or polarization propagators) \cite{usb09}, on the one hand, and that
of expressing such response functions in terms of slowly converging
sums over empty one-electron states \cite{Godby-Reining:1997,srr00,bg08,usb09},
on the other hand. In a recent paper, we have successfully addressed
the first problem by expressing polarizability operators in terms
of an optimally small set of basis functions \cite{usb09}. In the
present letter we address, and hopefully solve, the second problem
by introducing a new method to calculate polarizability operators
and self-energy operators, based on a Lanczos-chain technique, inspired
by recent progresses in time-dependent density-functional perturbation
theory \cite{tddfpt-1,tddfpt-2}. Our approach is first validated
by calculating the vertical ionization energies of the benzene molecule,
and its power demonstrated by addressing the spectrum of a large molecule
such as free-base tetraphenylporphyrin.

QP energies (QPE) are eigenvalues of a Schrödinger-like QP equation
(QPEq) for the so-called QP amplitudes (QPA), which is similar to
the DFT Kohn-Sham equation with the exchange-correlation potential,
$V_{xc}(\mathbf{r})$, replaced by the non-local, energy-dependent,
and non-Hermitian self-energy operator, $\Sigma(\mathbf{r},\mathbf{r}',\omega)$.
In the GWA \cite{h65,ag98} $\Sigma$ is the convolution of the one-electron
propagator, $G$, and of the dynamically screened interaction, $W$:
\begin{equation}
\Sigma_{GW}(\mathbf{r},\mathbf{r}';\omega)=\frac{i}{2\pi}\int_{-\infty}^{\infty}d\omega'G(\mathbf{r},\mathbf{r}';\omega')W(\mathbf{r},\mathbf{r}';\omega-\omega'),\label{eq:sigma-GW}\end{equation}
where $W=v+v\cdot\Pi\cdot v$, $\Pi(\mathbf{r},\mathbf{r}';\omega)=(1-P\cdot v)^{-1}\cdot P$
is the reducible polarizability, $P$ the irreducible one, $v(\mathbf{r},\mathbf{r}')=\frac{1}{|\mathbf{r}-\mathbf{r}'|}$
is the bare Coulomb interaction, and a dot indicates the product of
two kernels, such as in $v\cdot\Pi(\mathbf{r},\mathbf{r}',\omega)=\int d\mathbf{r}''v(\mathbf{r},\mathbf{r}'')\Pi(\mathbf{r}'',\mathbf{r}';\omega)$.
We assume time-reversal invariance to hold---so that wave-functions
are real---and we work on the imaginary-frequency axis \cite{rgn95}:
real-frequency results are then recovered upon analytic continuation.
One further approximation is the so called $\mathrm{G^{\circ}W^{\circ}}$
one, where the one-electron propagator is obtained from the QPEq using
a model real and energy-independent self-energy, such as \emph{e.g.
}$\Sigma^{\circ}=V_{{\rm xc}}(\mathbf{r})\delta(\mathbf{r}-\mathbf{r}')$,
and the irreducible polarizability is calculated in the random-phase
approximation (RPA): \begin{equation}
P^{\circ}(\mathbf{r},\mathbf{r}';i\omega)=4\mathrm{Re}\sum_{cv}\frac{\psi_{c}(\mathbf{r})\psi_{v}(\mathbf{r}')\psi_{v}(\mathbf{r})\psi_{c}(\mathbf{r}')}{i\omega-(\epsilon_{c}-\epsilon_{v})},\label{eq:rpa}\end{equation}
where $\psi$ and $\epsilon$ are zero-th order QPAs and QPEs, and
$v$ and $c$ suffixes indicate occupied and empty states, respectively.
To first order in $\hat{\Sigma}'=\hat{\Sigma}_{G^{\circ}W^{\circ}}-\hat{\Sigma}^{\circ}$,
QPEs are given by the equation: $E_{n}\approx\epsilon_{n}+\langle\hat{\Sigma}_{G^{\circ}W^{\circ}}(E_{n})\rangle_{n}-\langle\hat{V}_{XC}\rangle_{n},$
where $\langle\hat{A}\rangle_{n}=\langle\psi_{n}|\hat{A}|\psi_{n}\rangle$,
and quantum-mechanical operators are indicated by a caret.

In Ref. \cite{usb09}, we reported on a strategy to build an optimal
representation for the polarizability operators, in terms of a reduced,
yet controllable accurate, orthonormal basis set $\{\Phi_{\mu}(\mathbf{r})\}$:
\begin{equation}
P^{\circ}(\mathbf{r},\mathbf{r}',i\omega)=\sum_{\mu\nu}P_{\mu\nu}^{\circ}(i\omega)\Phi_{\mu}(\mathbf{r})\Phi_{\nu}(\mathbf{r}').\label{eq:Prepr}\end{equation}
Using the RPA, Eq. \eqref{eq:rpa}, $P_{\mu\nu}^{\circ}(i\omega)$
reads : \begin{multline}
P_{\mu\nu}^{\circ}(i\omega)=-4\mathrm{Re}\sum_{v,c}\frac{1}{\epsilon_{c}-\epsilon_{v}+i\omega}\times\\
\int d\mathbf{r}d\mathbf{r}'\Phi_{\mu}(\mathbf{r})\psi_{v}(\mathbf{r})\psi_{c}(\mathbf{r})\psi_{v}(\mathbf{r}')\psi_{c}(\mathbf{r}')\Phi_{\nu}(\mathbf{r}').\label{eq:P0matrix}\end{multline}

Suppose that such a representation for $P^{\circ}$ can be found without
any explicit reference to empty states (later we will show how this
can be achieved). In order to eliminate the sum over empty states
in Eq. \eqref{eq:P0matrix}, we introduce the projector operator over
the empty-state (\emph{electron}) manifold, $\hat{Q}_{e}=\hat{1}-\hat{Q}_{h}$,
$\hat{Q}_{h}$ being the projector onto occupied (\emph{hole}) states.
In terms of $\hat{Q}_{e}$, Eq. \eqref{eq:P0matrix} reads: \begin{multline}
P_{\mu\nu}^{\circ}(i\omega)=\\
-{\color{red}{\color{black}4}}\mathrm{Re}\sum_{v}\langle\psi_{v}\Phi_{\mu}|\hat{Q}_{e}(\hat{H}^{\circ}-\epsilon_{v}+i\omega)^{-1}\hat{Q}_{e}|\psi_{v}\Phi_{\nu}\rangle,\label{eq:polasenza}\end{multline}
where $\hat{H}^{\circ}$ is the QP Hamiltonian corresponding to $\hat{\Sigma}^{\circ}$
and $|\psi_{v}\Phi_{\nu}\rangle$ is the vector whose coordinate representation
is $\langle\mathbf{r}|\psi_{v}\Phi_{\nu}\rangle=\psi_{v}(\mathbf{r})\Phi_{\nu}(\mathbf{r}).$
A direct approach to Eq.\ \eqref{eq:polasenza} would require the
inversion of $(\hat{H}^{\circ}-\epsilon_{v}+i\omega)$ for every value
of the (imaginary) frequency and the application of the resulting
inverse to $N_{v}\times N_{P}$ vectors, where $N_{v}$ and $N_{P}$
are the number of valence states and polarizability basis functions,
respectively, thus making the resulting algorithm unwieldy. In order
to substantially reduce the computational load, we proceed in two
steps: we first reduce the number of functions to which the inverse
shifted Hamiltonian in Eq. \eqref{eq:polasenza} has to be applied;
in the second step, the inversion of the shifted Hamiltonian is avoided
by a Lanczos-chain algorithm that need not to be repeated for different
values of the (imaginary) frequency shift. In the first step an approximate
orthonormal basis, $\{t_{\alpha}(\mathbf{r})\}$, is built for the
linear space spanned by the $N_{v}\times N_{P}$ vectors, $\{\hat{Q}_{e}|\psi_{v}\Phi_{\nu}\rangle\}$:
\begin{equation}
\langle\mathbf{r}|\hat{Q}_{e}|\psi_{v}\Phi_{\mu}\rangle\approx\sum_{\alpha=1}^{N_{T}}t_{\alpha}(\mathbf{r})T_{\alpha,v\mu},\label{eq:span1}\end{equation}
where $T_{\alpha,v\mu}=\langle t_{\alpha}|\hat{Q}_{e}|\psi_{v}\Phi_{\mu}\rangle$
and $N_{T}$ is the number of $t$ functions, which can be kept in
general significantly smaller than $N_{v}\times N_{P}$. Details on
this procedure, based on a block version of the Gram-Schmidt algorithm,
will be given elsewhere. Here suffice it to say that for each $v$
index a block of $N_{P}$ vectors is first orthogonalized to the previously
processed block, and then reduced by eliminating the eigenvectors
of the overlap matrix corresponding to eigenvalues smaller than a
given threshold \cite{usb09}. Using Eq. \eqref{eq:span1}, Eq. \eqref{eq:polasenza}
reads:\begin{multline}
P_{\mu\nu}^{\circ}(i\omega)\approx-4\mathrm{Re}\sum_{v,\alpha\beta}\langle t_{\alpha}|(\hat{H}^{\circ}-\epsilon_{v}+i\omega)^{-1}|t_{\beta}\rangle\times\\
T_{\alpha,v\mu}T_{\beta,v\nu}.\label{eq:polase}\end{multline}

Having thus reduced the number of matrix elements of the resolvent
of the Hamiltonian in Eq. \eqref{eq:polasenza}, these matrix elements
can be efficientl\textcolor{black}{y calculated by a Lanczos-chain
algorithm \cite{lanczos}, as explained in Ref. \cite{tddfpt-2}.
In a nutshe}ll, a frequency-independent chain of vectors is generated
for each $t$ function. In the basis of each one of these chains,
the Hamiltonian is tridiagonal, and thus easily inverted. A similar
approach can be applied to the calculation of the expectation values
of the self-energy. By expressing the reducible polarizability in
the same representation used for the irreducible one, Eq. \eqref{eq:Prepr},
$\Pi(\mathbf{r},\mathbf{r}',\omega)=\sum_{\mu\nu}\Pi_{\mu\nu}(i\omega)\Phi_{\mu}(\mathbf{r})\Phi_{\nu}(\mathbf{r}')$,
the diagonal matrix elements of the correlation contribution $\hat{\Sigma}_{c}$
to the self-energy, Eq. \eqref{eq:sigma-GW}, can be cast into the
form: \begin{multline}
\langle\hat{\Sigma}_{c}(i\omega)\rangle_{n}=\frac{1}{2\pi}\sum_{\mu,\nu}\int d\omega'\Pi_{\mu\nu}(i\omega')\times\\
\langle\psi_{n}(v\Phi_{\mu})|(\hat{H}^{\circ}-i(\omega-\omega'))^{-1}|\psi_{n}(v\Phi_{\nu})\rangle,\label{eq:sigmashort}\end{multline}
where $|\psi_{n}(v\Phi_{\mu})\rangle$ is the vector whose coordinate
representation reads: $\langle\mathbf{r}|\psi_{n}(v\Phi_{\mu})\rangle=\psi_{n}(\mathbf{r})\int v(\mathbf{r},\mathbf{r}')\Phi_{\mu}(\mathbf{r}')d\mathbf{r}'.$
The matrix elements on the r.h.s. of Eq. \eqref{eq:sigmashort} are
calculated with a similar procedure as for Eq. \eqref{eq:polasenza},
where the set of vectors $\{|\psi_{n}(v\Phi_{\mu})\rangle\}$ is first
expanded into a suitable optimal basis set, $\{s_{\alpha}(\mathbf{r})\}$:
\begin{equation}
\langle\mathbf{r}|\psi_{n}(v\Phi_{\mu})\rangle\thickapprox\sum_{\alpha=1}^{N_{S}}s_{\alpha}(\mathbf{r})S_{\alpha,n\mu},\label{eq:span2}\end{equation}
 and then by generating a Lanczos chain for each $s$; the convolution
is finally calculated either by direct integration or by fast Fourier
transform. 

In Ref. \cite{usb09} a reduced basis set for the polarizability operators
was constructed by expressing the product functions, $\psi_{c}(\mathbf{r})\psi_{v}(\mathbf{r})$,
in Eq. \eqref{eq:rpa} in terms of localized Wannier-like orbitals.
Although the number of empty states needed to achieve a good accuracy
can be kept fairly small, still quite a few of them have to be calculated.
On the other hand, it was noted that this basis can be kept independent
on frequency (or on time). An optimal representation for the polarizability
can be thus calculated by diagonalizing the irreducible polarizability
operator at $t=0$ and keeping only those eigenvectors that correspond
to eigenvalues larger than a certain threshold. One has:\begin{align}
P_{\circ}(\mathbf{r},\mathbf{r}';t=0) & =\sum_{cv}\psi_{c}(\mathbf{r})\psi_{v}(\mathbf{r})\psi_{c}(\mathbf{r}')\psi_{v}(\mathbf{r}')\nonumber \\
 & =Q_{h}(\mathbf{r},\mathbf{r}')Q_{e}(\mathbf{r},\mathbf{r}'),\label{eq:P00}\end{align}
where $Q_{h}$ and $Q_{e}$ are real-space representations of the
projectors onto the hole and electron manifolds, respectively. The
eigenpairs of $P_{\circ}(t=0)$ can be easily calculated by iterative
diagonalization, noting that $Q_{e}(\mathbf{r},\mathbf{r}')=\delta(\mathbf{r}-\mathbf{r}')-Q_{h}(\mathbf{r},\mathbf{r}')$.
Such a procedure would lead however to a number of eigenpairs much
larger than strictly needed to achieve a good accuracy in the QP spectra.
In order to keep the size of the polarizability basis manageable,
we replace $\hat{Q}_{e}$ in Eq. \eqref{eq:P00} with the projector
over the manifold spanned by plane waves (PWs) up to a kinetic energy
of $E^{\star}$, orthogonalized to the hole manifold (\emph{orthogonalized
plane waves}, OPWs), $\hat{Q}_{e}^{\star}$. Let $|\bar{\mathbf{G}}\rangle=\hat{Q}_{e}|\mathbf{G}\rangle$
be one such OPW. In terms of the OPWs, the modified projector reads:

\begin{equation}
\hat{Q}_{e}^{\star}=\sum_{|\mathbf{G|^{2},|\mathbf{G'}|^{2}}\le E^{\star}}\bar{\mathbf{|G}}\rangle\langle\mathbf{\bar{G}}'|\times S_{\mathbf{G}\mathbf{G}'}^{-1},\label{eq:Q-star}\end{equation}
where $S_{\mathbf{G}\mathbf{G}'}=\langle\bar{\mathbf{G}}|\bar{\mathbf{G}}'\rangle$.
Using this approximation, a basis for the polarizability can be obtained
from the eigenfunctions of the eigenvalue equation:

\begin{equation}
\sum_{v}\psi_{v}(\mathbf{r})\langle\mathbf{r}|\hat{Q}_{e}^{\star}|\psi_{v}\Phi_{\mu}\rangle=q_{\mu}\Phi_{\mu}(\mathbf{r}),\label{eq:polabase2}\end{equation}
whose leading eigenpairs---corresponding to eigenvalues larger than
a given threshold, $q^{\star}$---can be easily found by conjugate
gradients or other iterative methods \cite{lanczos}. We stress that
this is a controlled approximation, which can be systematically improved;
furthermore, it only affects the determination of an optimal representation
for the polarizability operators, not their actual calculation once
this basis has been determined.

\textcolor{black}{}%
\begin{figure}
\textcolor{black}{\includegraphics[angle=-90,width=8cm]{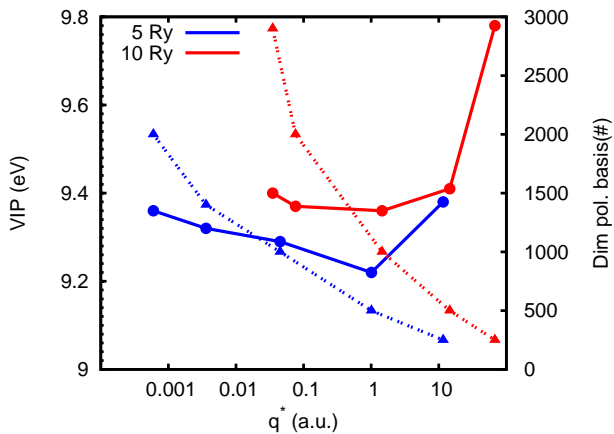}\caption{\label{fig:ip} Calculated vertical ionization potential of the benzene
molecule (discs, left scale) and dimension of the polarizability basis
(triangles, right scale) versus the $q^{*}$cutoff. The polarizability
basis sets have been constructed with energy cutoffs: $E^{*}=5\,\mathrm{Ry}$
(blue) and $E^{*}=10\,\mathrm{Ry}$  (red). The lines are a guide
to the eye.}
}

\textcolor{black}{\vskip-0.5cm}
\end{figure}
 \textcolor{black}{Our scheme has been implemented for norm-conserving
pseudopotentials (PPs) in the }\texttt{\textcolor{black}{gww.x}}\textcolor{black}{
\ module of the }\textsc{\textcolor{black}{Quantum ESPRESSO}}\textcolor{black}{
\ distribution of electronic-structure codes \cite{qe}, soon to
be released under the GPL license, and benchmarked on the isolated
benzene molecule \cite{technicalities}. In Fig.\ \ref{fig:ip} we
display the vertical ionization potential (VIP) calculated for the
isolated benzene ($\mathrm{C_{6}H_{6}}$) molecule with different
values of the energy and eigenvalue cutoffs, $E^{\star}$ and $q^{*}$,
defining the polarizability basis (see Eqs. \ref{eq:Q-star} and \ref{eq:polabase2}).
The two series of calculations for $E^{*}=5$ Ry and $E^{*}=10$ Ry
converge to the same VIP within few tens of meV . For both values
of $E^{\star}$, a cutoff $q^{*}\sim10\mathrm{\, a.u.}$ yields convergence
within $\sim0.1\,\mathrm{eV}$, which is our estimated residual accuracy
due to the incertitudes of the analytical continuation procedure.
The convergence of the VIP with respect to the size of the polarizability
basis is slightly slower here than previously observed in Ref.\ 
\cite{usb09}, where sums over empty states where performed explicitly
including a limited number of them. In Fig.\ \ref{fig1} we display
the differences between the calculated and experimental VIPs in benzene
($E^{\star}=10$ Ry, $q^{\star}=0.035$ a.u., corresponding to $\sim2900$
basis functions). On the same figure we also report results obtained:
}\textcolor{black}{\emph{i)\  }}\textcolor{black}{with a reduced
polarizability basis ($E^{\star}=10$ Ry, $q^{\star}=14.5.$ a.u.
corresponding to $\sim500$ basis functions); }\textcolor{black}{\emph{ii)\ }}\textcolor{black}{
with the extrapolation of the sum over virtual orbitals described
in Ref.\ \cite{usb09};}\textcolor{black}{\emph{ iii)\ }}\textcolor{black}{
with a polarizability basis of $\sim1500$ PWs (corresponding to a
kinetic-energy cutoff of $5\,\mathrm{Ry}$). Not unexpectedly, GW
results are in good agreement with experiment. What matters here is
that the present approach is not only considerably faster, but even
more accurate, than previous GW calculations that required extrapolation
of slowly converging sums over empty states \cite{usb09}. Moreover,
for a same size of the polarizability basis, the optimal polarizability
basis method is also more accurate than the use of simple PW basis
sets.}

\begin{figure}
\includegraphics[width=6cm]{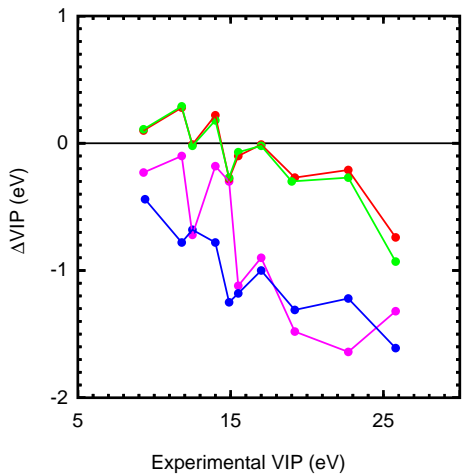} 

\caption{\label{fig1} Differences between the calculated VIPs of benzene and
experimental results. Re\textcolor{black}{d and green: present method
using $E^{\star}=10\,\mathrm{Ry}$, $q^{\star}=0.035\,\mathrm{a.u.}$
and $E^{\star}=10\,\mathrm{Ry}$, $q^{\star}=14.5\,\mathrm{a.u.}$
respectively.\ }\textcolor{red}{ }\textcolor{black}{Magenta: method
of Ref. \protect\cite{usb09}, upon extrapolation of the sum over
virtual states. Blue: present method, but using a polarizability basis
formed by PWs (see text). Experimental data are from Ref.\ \protect\cite{ldp76}.
The lines are a guide to the eye. }}

\end{figure}

\begin{figure}
\includegraphics[angle=-90,width=8cm]{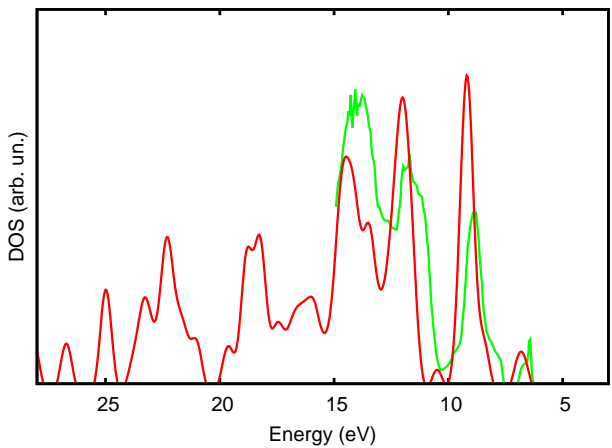} 

\caption{\label{fig2} Valence electronic density of states for the TPPH$_{2}$
molecule, red: theory (a Gaussian broadening of 0.25 eV has been used);
green: photoemission data from Ref.\ \protect\cite{glo99}. }

\vskip-0.5cm
\end{figure}

We now demonstrate the potential of our method by considering the
free-base tetraphenylporphyrin molecule (TPPH$_{2}$) ($\mathrm{C_{44}H_{30}N_{4}}$)
\cite{technicalities}. We used a polarizability basis of 5000 elements,
which has been obtained from Eq. \eqref{eq:polabase2} using $E^{\star}=10\,\mathrm{Ry}$
and $q^{\star}=21.1\,\mathrm{a.u.}$ These parameters ensure an absolute
convergence of the calculated QPEs of $\sim0.1\,\mathrm{eV}$. We
note that the analytic continuation procedure involves incertitudes
of similar size or even larger for the lowest lying states. The calculated
ionization potential is 6.7 eV, in fair agreement with an experimental
value of 6.4 eV \cite{glo99}. The quality of our results is further
illustrated in Fig.\ \ref{fig2}, where we compare the calculated
valence electronic density of states with photoemission data from
Ref.\ \cite{glo99}. \textcolor{black}{Nice agreement is achieved
for the position of the peaks, while the agreement for the intensities
is not as good, possibly due to our neglect of any matrix-element
effect.}

In conclusion, we believe that the method presented here may open
the way to \emph{ab-initio} MBPT simulations of large and realistic
models of molecular and nano-structured systems. While we feel that
the proposed Lanczos technique may be considered as a sort of definitive
answer to the sum-over-virtual-states problem, we think that there
is still room for improving the construction of an optimal polarizability
basis. The extension of these ideas to real-frequency implementations
of GW and other MBPT techniques, such as the Bethe-Salpeter equation,
is possible, and indeed presently under way.

P.U. thanks Xiaofeng Qian and Nicola Marzari for useful discussions
and the latter for hospitality at MIT.

\bibliographystyle{PhysRev}

\end{document}